\documentclass[twocolumn]{aastex631}
\usepackage{mathrsfs}

\shorttitle{Anisotropy of turbulent flows in flare ALT}
\shortauthors{Xie et al.}
\graphicspath{{./}{figures/}}
\begin{document}
\title{Anisotropic Turbulent Flows Observed in Above the Loop-top Regions During Solar Flares}

\author{Xiaoyan Xie}\thanks{E-mail: xiaoyan.xie@cfa.harvard.edu}
\affiliation{Harvard-Smithsonian Center for Astrophysics, 60 Garden Street, Cambridge, MA 02138, USA}
\author{Chengcai Shen}
\affiliation{Harvard-Smithsonian Center for Astrophysics, 60 Garden Street, Cambridge, MA 02138, USA}
\author{Katharine K. Reeves}
\affiliation{Harvard-Smithsonian Center for Astrophysics, 60 Garden Street, Cambridge, MA 02138, USA}
\author{Bin Chen}
\affiliation{Center for Solar Terrestrial Research, New Jersey Institute of Technology, University Heights, Newark, NJ 07102, USA}

\author{Xiaocan Li}
\affiliation{Los Alamos National Laboratory, Los Alamos, NM 87545, USA}

\author{Fan Guo}
\affiliation{Los Alamos National Laboratory, Los Alamos, NM 87545, USA}

\author{Sijie Yu}
\affiliation{Center for Solar Terrestrial Research, New Jersey Institute of Technology, University Heights, Newark, NJ 07102, USA}

\author{Yuqian Wei}
\affiliation{Center for Solar Terrestrial Research, New Jersey Institute of Technology, University Heights, Newark, NJ 07102, USA}

\author{Chuanfei Dong}
\affiliation{Center for Space Physics and Department of Astronomy, Boston University, Boston, MA 02215, USA}


\begin{abstract}

Solar flare above-the-loop-top (ALT) regions are vital for understanding solar eruptions and fundamental processes in plasma physics. Recent advances in 3D MHD simulations have revealed unprecedented details on turbulent flows and MHD instabilities in flare ALT regions. Here, for the first time, we examine the observable anisotropic properties of turbulent flows in ALT by applying a flow-tracking algorithm on narrow-band Extreme Ultraviolet (EUV) images that are observed from the face-on viewing perspective. First, the results quantitatively confirm the previous observation that vertical motions dominate and that the anisotropic flows are widely distributed in the entire ALT region with the contribution from both upflows and downflows. Second, the anisotropy shows height-dependent features, with the most substantial anisotropy appearing at a certain middle height in ALT, which agrees well with the MHD modeling results where turbulent flows are caused by Rayleigh–Taylor-type instabilities in the ALT region. Finally, our finding suggests that supra-arcade downflows (SADs), the most prominently visible dynamical structures in ALT regions, are only one aspect of turbulent flows. Among these turbulent flows, we also report the anti-sunward-moving underdense flows that might develop due to MHD instabilities, as suggested by previous three-dimensional flare models. Our results indicate that the entire flare fan displays group behavior of turbulent flows where the observational bright spikes and relatively dark SADs exhibit similar anisotropic characteristics.

\end{abstract}


\section{Introduction} \label{sec:intro}
Solar flares are one of the most energetic activities in the solar system \citep{2000JGR...10523153F}. The above the loop-top (ALT) regions that connect the reconnection current sheet and post flare loops are pivotal for understanding the evolution of solar eruptions \citep{2011ApJ...727L..52R, 2018ApJ...864...63P,2017A&A...604L...7M,2020ApJ...900...17Y,2022MNRAS.509..406X} and other physical processes, including particle acceleration \citep{2009A&A...494..669M,2015Sci...350.1238C,2020NatAs...4.1140C, 2020ApJ...905L..16K, 2022Natur.606..674F,2022ApJ...932...92L}, turbulence evolution \citep{2017PhRvL.118o5101K,2018ApJ...866...64C,2018ApJ...854..122W, Dong2018,Dong2022,2024FrASS..1183746X,2024ApJ...973...96A,2011ApJ...740...70M}, plasma instabilities \citep{2014ApJ...796L..29G,2014ApJ...796...27I,2022ApJ...931L..32W}, and shock waves and related plasma oscillations \citep{2016ApJ...823..150T,2017ApJ...848..102T,2018ApJ...869..116S,2020ApJ...905..165R,2022MNRAS.509..406X,2023ApJ...943..106S}. 

Flare supra-arcade fans are among the plethora of ALT features observed from a face-on viewing perspective (see the diagram in Figure~\ref{evolution}). These fans are full of dynamical structures and turbulent flows \citep{2013ApJ...766...39M,2014ApJ...788...26D,2014ApJ...796L..29G,2018ApJ...866...29F,2021A&A...653A..51L}. The most prominent dynamical structures in extreme-ultraviolet (EUV) and soft X-ray observations of flare fans are supra-arcade downflows (SADs, see the structures indicated by red arrows in Figure~\ref{evolution}(d), \citealp{1999ApJ...519L..93M,2000SoPh..195..381M,2003SoPh..217..247I,2009ApJ...697.1569M,2010ApJ...722..329S,2011ApJ...730...98S,2011ApJ...742...92W,2012ApJ...747L..40S,2014ApJ...796...27I,2017A&A...606A..84C}). SADs are underdense finger-shaped structures, with a typical width of approximately 5 Mm and speeds of about 100 km~s$^{-1}$, moving sunwards \citep{2011ApJ...730...98S,2022ApJ...933...15X}. SADs could contribute to plasma heating in the fan \citep{2017ApJ...836...55R,2020ApJ...898...88X,2021ApJ...915..124L,2023ApJ...942...28X} and have been thought to be correlated with nonthermal bursts of hard X-rays \citep{2004ApJ...605L..77A} and quasi-periodic pulsations (QPPs) in solar radiation \citep{2021Innov...200083S}. Another dynamic structure in ALT regions is the bright, spike-like features among dark SADs (see structures indicated by white arrows with annotation "s" in Figure~\ref{evolution}). Studies of \citet{2021A&A...653A..51L} show that the widths of these spikes follow the log-normal distributions, similar to those of SADs \citep{2011ApJ...735L...6M,2022ApJ...933...15X}, suggesting a plausible correlation of the formation of SADs and spikes in flare fans. 

Recent high-resolution 3D magnetohydrodynamic (MHD) simulations have revealed the presence of various MHD instabilities in ALT regions, providing insights into the origin and evolution mechanisms of various turbulent flows, such as SADs and spikes, within ALT regions. \citet{2022NatAs...6..317S} illustrated the formation of interface layers with turbulent flows below the termination shock where the reconnection outflows meet the flare arcades in the ALT region due to the non-linear development of Rayleigh–Taylor instability (RTI) and the Richtmyer–Meshkov instability (RMI). A group of turbulent flows, manifesting as underdense downflows, are in accordance with SADs in observations. \citet{2023ApJ...943..106S} revealed a rapid growth of MHD instabilities in the arms of the so-called magnetic tuning fork located at the upper parts of the ALT region and the generation of turbulent flows within the ALT region. 
3D simulations of \citet{2023ApJ...947...67R} demonstrate Kelvin-Helmholtz instability (KHI) can generate turbulent plasma motions in ALT regions due to the nonlinear interaction between the reconnection outflows and the flare arcades below the magnetic reconnection site. The turbulence is anisotropic due to magnetic tension, and it can spread as Alfv\'{e}nic perturbations and leads to a broad spatial distribution of turbulent motions.

In this paper, we investigate the properties of turbulent flows in the supra-arcade fan of a newly erupted solar flare (see Figure~\ref{evolution}). Following the approach in \citet{2013ApJ...766...39M}, we apply a flow tracking algorithm to facilitate the characterization of turbulent flows. \citet{2022NatAs...6..317S} illustrated that the turbulent flows favorably form in the areas where $\beta$ (the ratio of thermal pressure to magnetic pressure) $\simeq$ 1, suggesting turbulent flows in flare ALT have certain variability with height. The properties of turbulent flows at different heights will be explored in this work. As supported by \citet{2013ApJ...766...39M} and \citet{2014ApJ...788...26D}, we reveal the anisotropic properties of these ALT turbulent flows quantitatively from observations, which compare favorably with the 3D modeling results. We introduce observations and tracking algorithms in the next section. The results and discussion are given in Sections~\ref{results} and~\ref{discussion}, respectively.
. 
\section{Observations \& Measurements}
\label{method}

\begin{figure*}[ht!]
\centering
\includegraphics[scale=0.8]{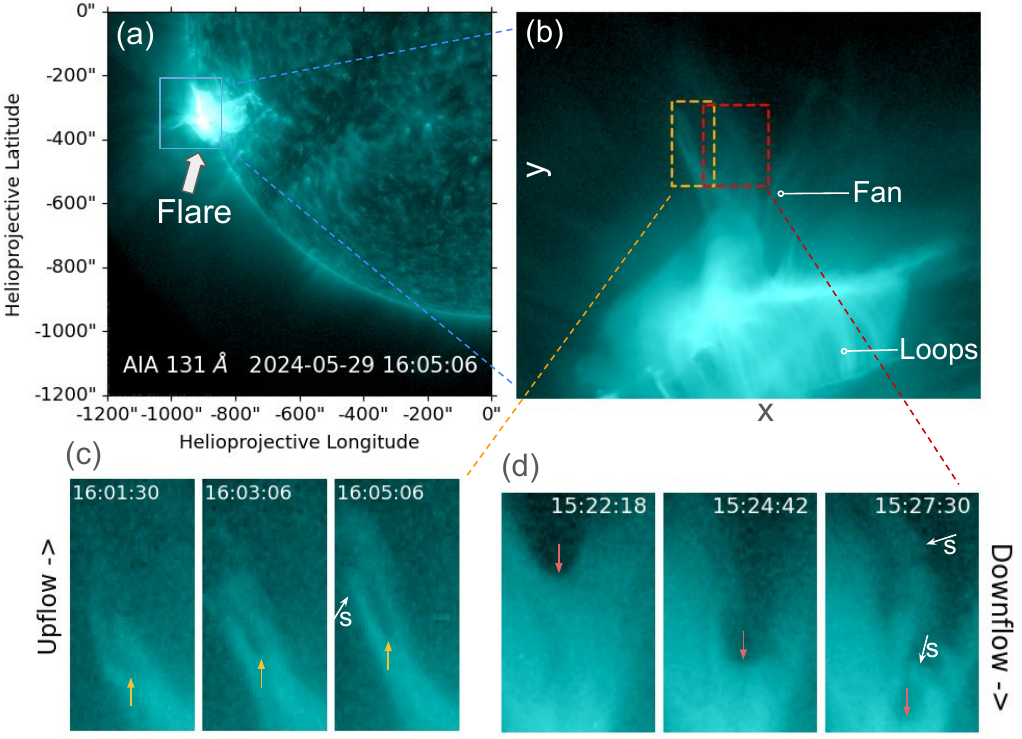}
\caption{A snapshot of the flare where the loops and supra-arcade fan are annotated. Panel (b) is the zoom-in of the blue box in panel (a), and we make a 90$^o$-clockwise rotation from the original AIA observation and define $x$ as the direction parallel to the flare loop-top. Downwards/upwards is in respect to the direction moving sunwards/anti-sunwards. An animation of panel (b) for 15:19:54 to 16:31:54 UT is available. Panel (c) displays the evolution of upflows corresponding to a zoom-in of the yellow box. We note the split of fan plasma becomes more pronounced over time. The yellow arrows mark the motion of the upward-moving underdense structure. Panel (d) displays the evolution of downflows corresponding to a zoom-in of the red box. The red arrows mark the motion of the SAD. The white arrows with annotation ``s" mark some of the spikes.}
\label{evolution}
\end{figure*}

On 2024 May 29, a flare of magnitude X1 erupted on the southeast of the Sun. It started at 14:11 UT, and the X-ray intensity peaked at 14:37 UT.  During the eruptions, flare loops form, and supra-arcade fan forms above the loops. SDO/AIA captures the flare in the EUV channel with a pixel size of 0.6$^{"}$ (435 km) and a cadence of 12 s. The turbulent flows are observed during the formation of the fan (see the online animation of Figure~\ref{evolution}), and the fan is fragmented with intensity-depleted structures (see Figure~\ref{evolution}) and spikes (some of the spikes are marked by white arrows with annotation ``s" in Figure~\ref{evolution}) in between. For the study of the paper, we make a 90$^{\circ}$ clockwise rotation from the original helioprojective coordinates and define $x$ as the direction nearly parallel to the loop-top in the fan plane and $y$ is perpendicular to the loop-top (Figure~\ref{evolution}(b)). We define the flows moving sunwards as downflows and anti-sunwards as upflows. An example of the evolution of a SAD is indicated by red arrows in Figure~\ref{evolution}(d). The SAD moves with a speed of $\sim$ 100 km/s and shrinks over time. The fan is split as SADs that penetrate into the fan. When multiple SADs are captured in the field of view (FOV) simultaneously, the same fan areas have a branch shape (e.g., the right panel in Figure~\ref{evolution}(d)).

The split of the fan is also shown in Figure~\ref{evolution}(c), where we note more bifurcating effects of the fan over time. Both the spikes and the low-intensity plasma in-between move upwards. The upward-moving plasma flows above flare arcades have been observed in \citet{2014ApJ...785..106K}, \citet{2019PhRvL.123c5102S} and \citet{2021Innov...200083S}. Density diagnostics indicate that the low-intensity structures in between the spikes in Panels (c) and (d) are both density-depleted (see Figure~\ref{ncurve}).  We have more than one under-intensity upflow shown in this event (see the movies ``upflow0.mp4" and ``upflow1.mp4"). Note that the simple loop shrinking and plasma accumulation from the site of magnetic reconnection to the flare loop-top are not supposed to have the underdense structures moving upwards. Instead, the simultaneous upward-moving underdense structure and the split of bright fan plasma have been illustrated in the modeling of \citet{2022NatAs...6..317S} where underdense structures in ALT are formed through the mechanisms of RTI and RMI. 

For the rest of this paper, we do not separate the upward/downward or brighter/dark flows in the fan. Instead, we treat them as the observational manifestation of turbulent flows in the ALT region and investigate their properties as a whole. As the readers will see in the following text, such a treatment can help examine if different types of turbulent flows come from the same origin of physical processes. We focus on the group behavior of turbulent flows and the corresponding consequences on the formation of the fan in ALT regions. To achieve this goal, we apply the Farnebäck flow-tracking algorithm \citep{10.1007/3-540-45103-X_50}, which estimates the motion of the object between two frames based on polynomial expansion and calculates the motion of each pixel in the FOV, on AIA 131 \AA\ to track flows in the fans. The code of Farnebäck’s algorithm is available in the Open Source Computer Vision Library \footnote{https://github.com/opencv/} (OpenCV), and the major procedures of performing the algorithm include polynomial expansion transform (for approximating the neighborhood), displacement estimation, and the refinements. The algorithm has been successfully applied to the solar activities \citep{2019A&A...624A..72Z} and exhibited good performance in capturing variant velocities of plasma motions in fans  \citep{2022ApJ...933...15X,2023ApJ...942...28X}. Considering the pixel size and time cadence of the data we use are 435~km and 24~s, respectively, we only consider the motion of the objects with speeds higher than 18 km/s. As shown in Figure~\ref{evolution}, the flare loop-top is almost horizontal to $x$-axis after we rotate 90 degrees clockwise. Therefore, we use $v_x$ and $v_y$ to indicate the movements horizontal and perpendicular to the loop-top, respectively. While the tracked velocities are the plane-of-sky velocity, we inevitably have some projection effects that come from the supra-arcade fan tilting toward/away in the 3rd dimension (the direction perpendicular to the plane of the sky). However, the projection effects from 3rd dimension equally scale the $v_x$ and $v_y$ from tracked results, it would not have an impact on the objective, anisotropic properties of flare fan, of the paper.

\section{Results} 
\label{results}
\begin{figure*}[ht!]
\centering
\includegraphics[scale=0.76]{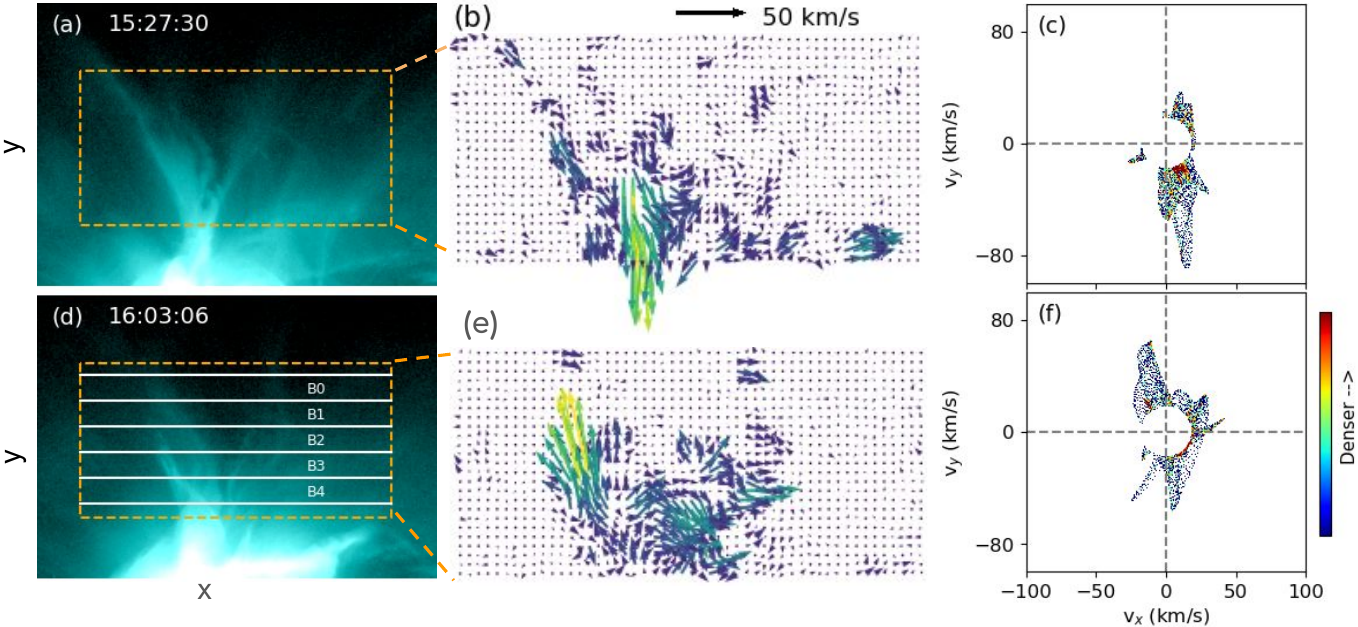}
\caption{AIA 131 ~\AA\ observations for fan at 15:27:30 (a) and 16:03:06 (d). The arrows in panels (b) and (e) are the corresponding velocity vectors with a more yellow color indicating higher speed. Panels (c) and (f) are $v_y$ versus $v_x$ (we call it $v$-map hereafter), where each dot indicates one pixel at 15:27:30 and 16:03:06, respectively. The white disk corresponds to the speed of plasma motions $<$ 18 km/s. Color towards red indicates denser distribution in $v$-map. B$_0$ to B$_4$ in panel (d) indicates the areas for checking $v$-map properties at different heights of the fan in Figure~\ref{bands}.}
\label{vmap}
\end{figure*}
Figure~\ref{vmap} displays the flow tracking results at two selected snapshots at 15:27:30 UT (panels (a)-(c)) and 16:03:06 UT (panels d-f), in which the arrows indicate velocity vectors. As shown in panels (a)-(b), the most prominent movements at this time are SADs (see the movie of ``SADs.mp4" or the evolution of some SADs in Figure~\ref{evolution}d), and yellow arrows indicate the highest speed of plasma. In addition to SADs, other upward- and downward-moving turbulent flows are captured by the flow tracking algorithm as well. We further plot the tracked flow velocity distribution, $v_y$ versus $v_x$ (referred to as the $v$-map hereafter), in Figure~\ref{vmap}(c) for the plasma motion of each pixel. The shapes of $v$-maps are different and represent velocity fields at different moments. We note that in Figures~\ref{vmap}(b) and (c), when downflows are dominant at the moment, $v$-map is elongated along negative $v_y$ direction. In comparison, the flow is dominated by both upward and downward flows present at 16:03:06 UT, as shown in Figure ~\ref{vmap}(d)-(f) at 16:03:06 UT (also see the movie of ``upflows0.mp4" for the evolution of prominent upflows). This exhibits an elongated $v$-map along both the positive and negative $y$ directions, forming a shape for this moment. In either case, the $v$-maps clearly demonstrate the anisotropic nature of turbulent flows in the fan area with $v_y \gg v_x$, and the major slope elongates the direction perpendicular to the solar surface as the fan is dominated by flows moving either upward or downward.
\begin{figure*}
\centering
\includegraphics[scale=0.52]{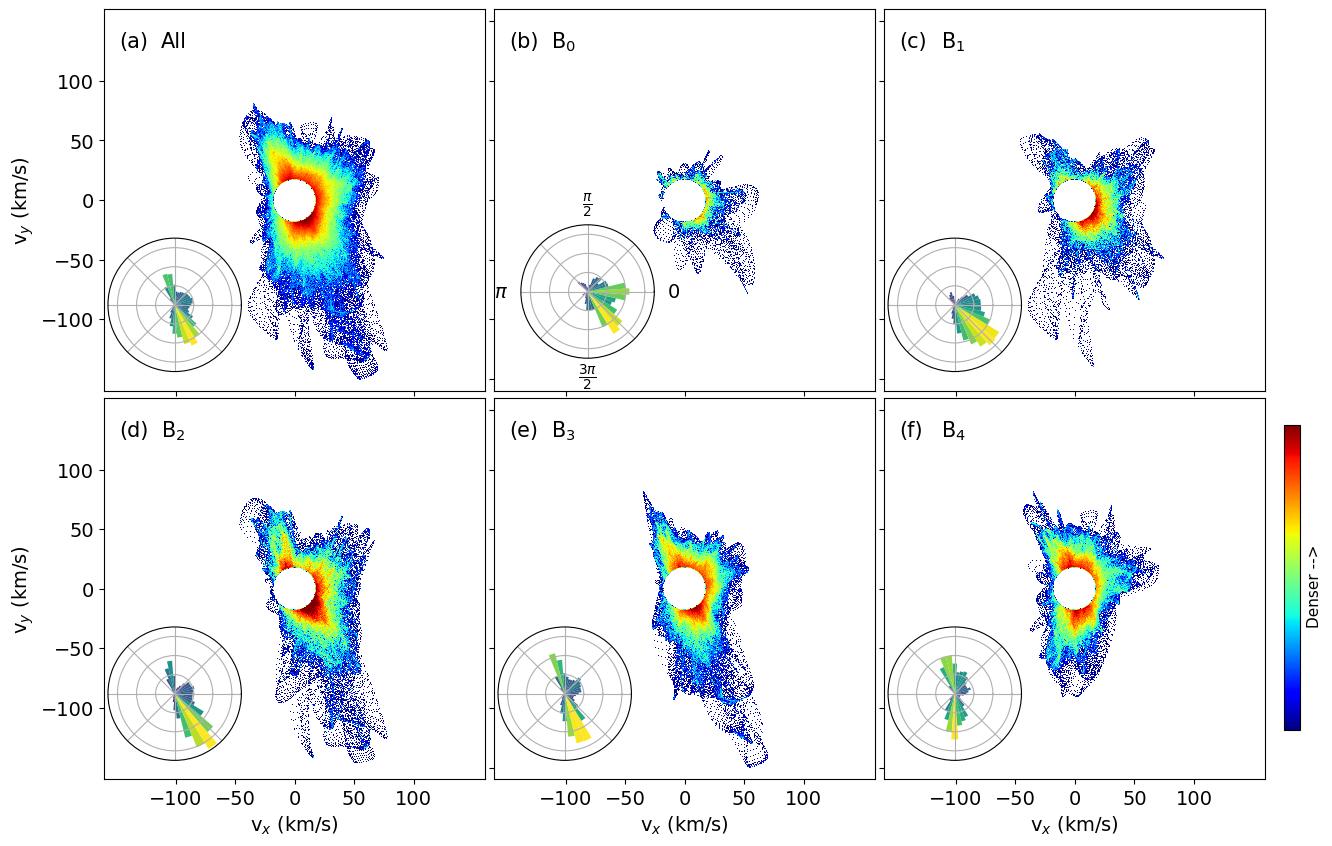}
\caption{$V$-map for the whole fan area (a) and different heights (b)-(f) indicated by B$_0$ to B$_4$ in Figure~\ref{vmap} where the white disk corresponds to the speed of plasma motions $<$ 18 km/s. The time period for $v$-maps is from 15:19:54 UT to 16:08:42 UT during the flare decay phase with the formation of the supra-arcade fan and the manifestations of upflows and downflows traveling through the fan. The histogram (probability density) of $\theta$ ($\theta$ is calculated by $\arctan(\frac{v_y}{v_x}$) and take the direction of upward and downward moving flows into account within the range of 0-2$\pi$) for the $v$-map is included in each panel. The gray circles moving outwards indicate probability of 0.2, 0.4, and 0.6, respectively.}
\label{bands}
\end{figure*}

\begin{table*}
  \centering
    \begin{tabular}{cccccc}     
\hline
Area &  $\% (\frac{\pi}{4} < \theta < \frac{3\pi}{4}~or~\frac{5\pi}{4} < \theta \leq \frac{7\pi}{4}) $  & $\% ( \frac{\pi}{4} < \theta \leq \frac{3\pi}{4})$ &  $\% (\frac{5\pi}{4} < \theta \leq \frac{7\pi}{4})$     & Peak $\theta$   \\
\hline
All &  68.5 & 29.9 & 38.6   & 1.65$\pi$  \\  
B$_0$ &  44.9 & 16.4 & 28.5    & 1.7$\pi$ \\   
B$_1$ &  46.0 & 12.0 & 34.0   &  1.75$\pi$  \\ 
B$_2$ &  60.1 & 22.8 & 37.3    & 1.7$\pi$  \\ 
B$_3$ &  75.1 & 34.0 & 41.1    &  1.65$\pi$  \\  
B$_4$& 81.2 &  41.0  & 40.2  &    1.5$\pi$   \\ 
\hline
    \end{tabular}
\caption{{The properties of the turbulent flows at different heights of the fan. $\%$ indicates the percentage of turbulent flows with certain conditions. Peak $\theta$ corresponds to the highest probability density in the histograms in Figure~\ref{bands}}.}
  \label{table}
\end{table*}

To obtain general information about anisotropic properties during fan formation, we further analyze all turbulent flows from 15:19:54 UT to 16:08:42 UT with 24s time cadence to show the statistical characteristics of the fan. This selected time period is during the flare decay phase, featuring the formation of a clear supra-arcade fan on SDO/AIA 131 \AA\ band observations and the manifestations of upflows and downflows traveling through the fan. The corresponding $v$-map is shown in Figure~\ref{bands}. The global picture of the fan during the examination shows anisotropic properties as we note the shape of $v$-map elongates roughly along $y$-direction. The anisotropic distribution suggests that the turbulent flows are dominated by either upward- or downward-flow features, consistent with well-recognized macroscopic flowing structures, such as downwards-moving SADs and upwards-moving spikes. Furthermore, this $v$-map also reveals that the observed anisotropy appears across a wide velocity range, consistent with the typical SADs and spikes structures \citep{2011ApJ...730...98S,2021ApJ...915..124L,2022ApJ...933...15X,2022MNRAS.516.3120T}.

To quantitatively analyze the properties of the $v$-map, we obtained the histogram of velocity direction (denoted as $\theta$) in polar coordinates for all samples in the $v$-map, as shown by the insets in Figure~\ref{bands}. The corresponding parameters are also listed in Table~\ref{table}. Here, $\theta$ is calculated by $\arctan(\frac{v_y}{v_x}$) and takes the direction of the upward and downward-moving flows into account in the range of 0 to 2$\pi$ (see annotation in Figure~\ref{bands}(b)).
We aim to quantify how the flow direction concentrates in the vertical direction (the $y$ direction) with $\frac{\pi}{4} < \theta < \frac{3\pi}{4}$ or $\frac{5\pi}{4} < \theta < \frac{7\pi}{4}$. As shown in Figure~\ref{bands}(a) and the first row in Table~\ref{table}, there are $\sim 68.5\%$ of turbulent flows in this range. The histogram also reveals the peak $\theta$, the highest probability density, telling the direction that has the highest distribution in $v$-map.  Here, peak $\theta$ = 1.65$\pi$ for all sampling points indicates that the shape elongating $v$-map mostly originates from the top left to the bottom right, which is not exactly along the selected $y$-direction. This deviation from the $y$-direction might be affected by guide fields and shearing of the entire flare loop system in a long-duration evolution. Also the selection of the coordinate system itself - i.e., the $y$-direction is not exactly aligned with the current sheet at a given time. Therefore, we do not expect the same peak angle for the flare events.

We analyze the anisotropy properties at different heights that are illustrated by the horizontal white bands ($B_0$ to $B_4$) in Figure~\ref{vmap}. The corresponding $v$-maps and histogram profiles in polar coordinates at these bands are displayed in Figures~\ref{bands}(b)-(f). The highest speed of both downward and upward flows increases prominently from B0 to B3 and decreases from B3 to B4, which is consistent with the previous observations of SADs (see Figure~13 of \citet{2022ApJ...933...15X}) where the highest speed of SADs generally occurs around the middle height of the fan. As the height decreases from B0 to B4, the flow anisotropy apparently increases. The probability density of the histogram concentrates more in $\frac{\pi}{4} < \theta < \frac{3\pi}{4}$ and $\frac{5\pi}{4} < \theta < \frac{7\pi}{4}$ with the decrease of height. $\% (\frac{\pi}{4} < \theta < \frac{3\pi}{4})$ in Table~\ref{table} shows that flows appear less anisotropic in high altitudes (e.g., bands B0 and B1), and the probability density around $y$-direction increases from 44.9\% at B0 to 81.2\% at B4 (Table~\ref{table}). We also notice that the corresponding tilt of $v$-map shape becomes gradually more vertical from B1 to B4, as the peak $\theta$ decreases from 1.7$\pi$ at B0 and B1 to 1.5$\pi$ at B4 (Table~\ref{table}). The speed range for upward and downward flows changes simultaneously as a function of height, suggesting a plausible correlation of upward- and downward-moving flows in the $v$-map. In fact, the upflow components contribute significantly to the $v$-map with a magnitude comparable to those of the downflows, at some heights in flare ALT as demonstrated by the histograms in Figure~\ref{bands}, suggesting that the SADs (the most visible downflows) are embedded in the turbulent flows areas where both upward and downward moving flows are present. 

\begin{figure*}
\centering
\includegraphics[scale=0.85]{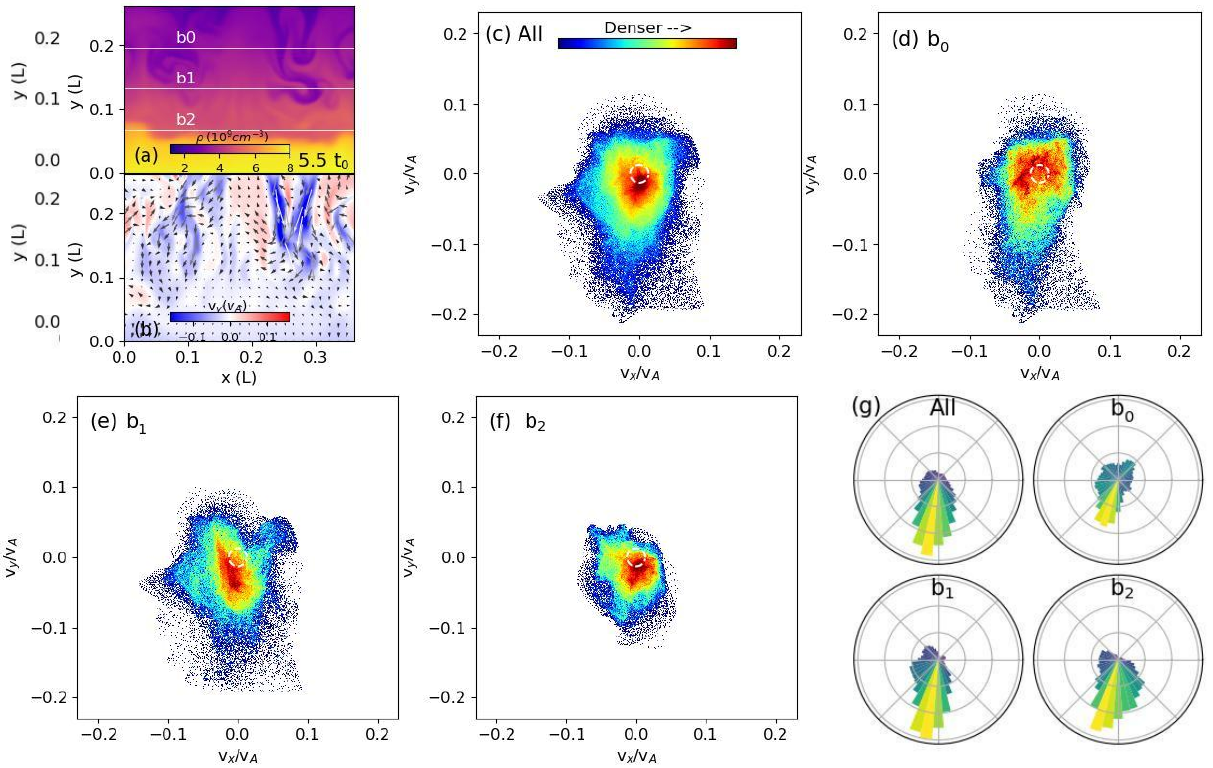}
\caption{Density distribution (a), $v_y$ distribution overlaid with velocity vectors (b), $v$-maps for the whole  area (c) and different heights (d)-(f) from the MHD simulation with the same dataset as in \citet{2022NatAs...6..317S}. Panels (a) and (b) are at time t=\textbf{5.5}t$_0$, and panels (c)-(g) are accumulated from t=\textbf{5.5}t$_0$ to t=7t$_0$ during which turbulent flows have been developed in the system. The corresponding height areas in (d)-(g) are indicated by b$_0$ to b$_3$ in (a). The corresponding histograms with the same direction definition as in Figure~\ref{bands} for $v$-maps (c)-(f) are shown in (g). The gray circles moving outwards indicate probability of 0.2, 0.4, \textbf{and 0.6}, respectively. Same as in the observations in the current paper, we define $x$ as the direction parallel to the loop-top, $x-y$ plane is the fan area, and $\theta$ is calculated by $\arctan(\frac{v_y}{v_x})$ within the range of 0-$\pi$. The simulation uses non-dimensional units with the characteristic parameters L=1.5$\times 10^{10}$ cm, V$_A$=1380 km/s, t$_0$=108.7 s. White dashed curve in panel marks v=$\sqrt{v_x^2+v_y^2}$=18 km/s. $\% (\frac{\pi}{4} < \theta < \frac{3\pi}{4}~or~\frac{5\pi}{4} < \theta \leq \frac{7\pi}{4}) $ of All, b$_0$, b$_1$, and b$_2$ is 67.1, 57.2, 65.2, and 63.0, respectively.} 
\label{simu}
\end{figure*}
One crucial question is whether the anisotropy properties described above can represent the nature of turbulent flows, given the observational EUV images are from optically thin emission signals, which are unavoidable to be affected by the projection effects and line-of-sight integration. Because the direct measurements of plasma velocity are currently unavailable, we employ a high-resolution numerical model to examine if these anisotropy characteristics match each other. 
We follow a comprehensive three-dimensional MHD simulation of solar flares reported by \citet{2022NatAs...6..317S}, in which the turbulent flows naturally developed in an interface region above post-flare loops initially driven by macroscopic MHD instabilities (e.g., RTI and RMI), which closely resemble the observational characteristics of SADs.
Thus, this numerical model can offer detailed information about the dynamics of these SAD-like structures in a 3D simulation domain in the ALT region.

Figures~\ref{simu}(a)-(b) shows the modeling density and plasma velocity ($v_y$) distribution overlaid with velocity vectors on a selected center plane. Here, the simulation parameters and datasets are the same as in \citet{2022NatAs...6..317S}.
Similar to the analysis methods on SDO/AIA images, we obtained the $v$-map and the corresponding histograms for the whole fan area and the different heights  of turbulent flows by using the simulation results, as displayed by Figures~\ref{simu}(c)-(g). Similar to the observations, we set $x$ as the direction parallel to the flare loop-top, and the $x-y$ plane is the fan area. We see that similar to observations, the shape of the $v$-maps for the whole fan and different heights elongate along the $y$-axis with $\% (\frac{\pi}{4} < \theta < \frac{3\pi}{4}~or~\frac{5\pi}{4} < \theta \leq \frac{7\pi}{4}) $ of All, b$_0$, b$_1$, and b$_2$ being 67.1, 57.2, 65.2, and 63.0, respectively, suggesting the presence of a similar anisotropy in the formation of the flare fans. We also plot the limit of applying tracking algorithms in our observations--18 km/s (indicated by white dashed circles in Figure~\ref{simu}) for reference. Both upward and downward-moving turbulent flows contribute to the shape of $v$-map, which is also reflected on the velocity field as the instability interface is filled with upward- and downward-moving turbulent flows simultaneously in this model (e.g., Figure ~\ref{simu}(b)). Similar to the observations, the speed range for upward and downward flows changes simultaneously as a function of height in the simulations. In addition, Figures~\ref{simu}(a)-(b) demonstrate that the highest value of $v_y$ occurs at certain heights of the fan where these instabilities can be activated relatively easily, which is consistent with the observations in Figure~\ref{bands} that shows the highest $v_y$ occurs in the certain middle heights- B$_2$ and B$_3$. 
The comparison with MHD models suggests that turbulent flows can be effectively characterized by their observable anisotropic properties, which are commonly observed in flare fan regions. These anisotropic properties are also reflected in other examined flare fan events (see Section~\ref{sec:otherevents} in the Appendix).

\section{Conclusion and Discussion} 
\label{discussion}
The ALT in solar flares is a highly dynamic region characterized by complex turbulent flows, as indicated by recent theoretical and observational studies. Prior high-resolution 3D magnetohydrodynamic (MHD) simulations have revealed the diverse structures within this region that are full of various MHD instabilities and turbulent flows occurring at different scales  (e.g.,\citealp{2022NatAs...6..317S,2023ApJ...943..106S,2023ApJ...947...67R,2023ApJ...955...88Y}). Specifically, \cite{2022NatAs...6..317S} demonstrated that, within the turbulent ALT region, both upward- and downward-moving flows are present to form SADs, as opposed to the dominating downward flows as suggested by certain models. On the other hand, the complex plasma flow signals are also reported from narrow-band high-temperature observations \citep{2013ApJ...766...39M,2018ApJ...866...29F}.
These findings indicate that complex plasma flows in ALT regions might play a critical role in energy transfer during solar flares. However, there is a lack of comprehensive understanding of these turbulent flows in the ALT regions.

In this work, we apply a flow tracking algorithm to identify these plasma flows that are observed from a face-on viewing angle in an extensive supra-arcade fan region. 
We examined their velocity distributions ($v$-map) over a long duration, analyzed their variation at different heights, and obtained the shape of $v$-map elongates along the direction perpendicular to the flare loop-top. Our results quantitatively confirm the previously qualitative observation \citep{2013ApJ...766...39M,2014ApJ...788...26D} that the anisotropy of turbulent flows appears in the entire flare ALT region (Table \ref{table}). Turbulent flows that move both upwards and downwards with speed smaller than 200 km/s contribute to the anisotropic property. The level of the anisotropy increases for upward and downward flows simultaneously with the decrease of the height from the loop-top. It changes from near isotropy at high height to pronounced anisotropy at a lower altitude close to the flare loop-top. 

It is worth noting that the flow-tracking velocities measured in our work differ from non-thermal velocities derived from spectra in supra-arcade fan (i.e., \citealp{2014ApJ...788...26D}). It is commonly thought that the non-thermal velocities come from the effects of turbulence and bulk flows along the line of sight (LOS). However, the derived non-thermal speed from spectral broadening can not offer information about the flow direction of those turbulence that appear in different length scales, as the spectral broadening is from the integrated effect contributed from flow components along LOS. In other words, the spectra analysis derived turbulent speed is only for these turbulent motions projected onto the chosen LOS. The velocities of turbulent flows from the flow tracking, on the other hand, give the information of plasma motions on the two-dimensional plane of sky. Thus, this approach allows us to explore the anisotropic properties of turbulent flows, although it is inevitably limited by the current spatial resolution of SDO/AIA. The correlation between the spectral-derived non-thermal velocities and those obtained by using flow-tracking is still unknown. Further 3D MHD simulations of solar flares with higher spatial resolutions would be beneficial to disentangle turbulent flows on multiple scales and from different viewing angles. Thus, the relevant comparison between the modeling predictions and the observational studies of both non-thermal velocities derived from spectra analysis and plasma motions from flow-tracking can provide more insights into the correlations between these two types of velocities and the nature of turbulent flows.

Compared with the numerical model of a solar flare, similar anisotropic features on turbulent flow velocity have been confirmed as well. Specifically, the anisotropic features due to the SADs-like turbulent flows revealed in the numerical model \citep{2022NatAs...6..317S} exhibit the characteristics of anisotropy shown in our observations. In this model, the RTI and RMI in an interface region above the flare loops are mainly developed along the solar radius direction, resulting in the turbulent flows caused by RTI and RMI favoring $y$ direction and forming the $v$-map elongating along $y$ direction, as shown in Figure~\ref{simu}. Also, the highest upward/downward-moving speeds occur in the middle height of the flare fan region, consistent with our findings from observations in Figure~\ref{bands}. As suggested by the model \citep{2022NatAs...6..317S}, the turbulent interface layer is generally associated with the condition of $\beta \simeq$ 1. In this scenario, both magnetic fields and plasma gas could significantly affect the system's evolution, creating favorable environments for various types of turbulent flows.
This condition is also consistent with the observational studies of turbulent flows \citet{2013ApJ...766...39M}, in which researchers showed plasma $\beta$ might exceed unity in flare fans.

It is worth mentioning that the above anisotropic characteristics are obtained based on all detectable plasma moving signals without specifically selecting well-identified SADs or the brightest spike structures. As shown in Figure~\ref{band-i} of Appendix, both the darker and brighter turbulent flows commonly display very similar anisotropic properties, suggesting the same origin as turbulent flows in flare fans. Our results serve as strong observational evidence supporting the model of the formation of SADs-like structures as proposed by \citet{2022NatAs...6..317S}, in which the authors illustrated that both high-density spikes and underdense bubbles are \textbf{naturally} developed structures due to RTI/RMI instabilities in a turbulent ALT region.
In this scenario, both bright (higher emission due to high density) and dark turbulent flows should display similar statistical features about their moving speeds on $v$-maps. Remarkably, in this event, we observed the clear upward dark flows on SDO/AIA images, which might be associated with the model predicted underdense upflows (see the deduced density maps in Figure~\ref{ncurve} of Appendix). Although occasional upward-moving flows have been reported in the literature, this is the first report, to our knowledge, to identify the ubiquitousness of the upflows in solar flare ALT fan regions alongside the well-known downflows or SADs. Both upflows and downflows contribute to anisotropic characteristics in the flare ALT regions, suggesting that downward-moving SADs and upward-moving spikes with various movement directions can be thought of as specific observational manifestations of more broadly existing turbulent flows in ALT regions. 

The mechanism behind these turbulent flows with various scales in ALT regions is still uncertain, though a set of recent numerical simulations suggested that the MHD instabilities can drive them. Recently, \citet{2023ApJ...955...88Y} showed that different anisotropy can be commonly found in magnetic reconnection regions, including buffer zones below the CMEs and large-scale magnetic reconnection current sheets. By performing 3D Fourier transform analysis on the velocity field obtained from a 3D MHD stimulation of solar CME/flare eruptions, they found that spectra distributions of the turbulent energy and anisotropy exhibited significant variation with height, which can significantly affect the energy release process.
In ALT regions, as discussed in the above sections (and illustrated in Figure~\ref{bands}), the varying anisotropic characteristics at different heights might indicate different turbulence mechanisms as well, potentially affecting the magnetic energy release and transport. In the middle height of flare fan regions, these large turbulent flows have been proposed to be relative to the non-linear developments of several MHD instabilities (e.g., RTI/RMI illustrated in \citealp{2022NatAs...6..317S} and KHI reported by \citealp{2023ApJ...947...67R}).
The turbulence can also form in the upper parts of the ALT region, called magnetic tuning forks regions, as reported by \citet{2023ApJ...943..106S}.
However, correctly revealing turbulent flows across different locations and scales in self-consistent CME/flare modeling remains challenging.
Therefore, our \textbf{quantitative} studies on these anisotropic features provide new insights into the nature of turbulence and raise exact requirements for future model development to explain this observable anisotropy.

In the literature, energy conversion and transport due to turbulence have been widely studied. For example, \citet{2023ApJ...947...67R} demonstrated that in ALT areas where KHI dominates with the formation of turbulent flows, the conversion from downflow kinetic energy into turbulence energy is more than 10$\%$. 
Thus, it is important to examine how the anisotropic turbulent flows affect the energy transportation process in ALT areas. 
On the other hand, the anisotropy of turbulence in ALT might potentially impact the transport of energetic particles during solar flares. 
In recent macroscopic MHD and particle transport modeling \citep[e.g.,][]{Kong2022ApJ...941L..22K, Chen2024ApJ...971...85C}, it has been found that the energetic electrons can be trapped in a magnetic bottle region located above the flare loop-top regions, due to turbulence (which is prescribed in their model). However, the understanding of the inherent properties of turbulence that govern particle transport equations remains challenging \citep[see details in][]{Li2019ApJ...884..118L}.  
Therefore, further investigations regarding the origin of turbulent flows, the anisotropic characteristics of turbulence in different areas in ALT, the role of turbulent flows in energy conversion, and the local plasma conditions revealed by the manifestations of turbulent flows are required to understand solar eruptions further.

Due to the resolution limitation of SDO/AIA, our current studies are unable to resolve even slower turbulent flows at their characteristic scales (e.g., 18 km/s). However, the properties of turbulent flows in small ranges are essential to understanding energy transport. In fact, \citet{2023ApJ...943..106S} have shown that ALT regions are full of turbulent flows from edge-on viewing perspective. Future instruments with higher resolution and cadence are required to further investigate the turbulent flows from different perspectives and piece together the 3D picture of more precise properties of turbulent flows in ALT. 

\begin{acknowledgments}
We  thank the anonymous referee for comments that improved the paper. X.X., C. S., and K. R. are supported by NSF grant AGS-2334929. B.C., S.Y., and Y.W. acknowledge the support from NSF grants AST-2108853, AGS-2334931, and NASA grants 80NSSC24K1242 and NNH240B72A. C.D. is supported by NSF grant AGS-2301338, DOE grant DE-SC0024639, and the Alfred P. Sloan Foundation. X.L. acknowledges the support from NASA through Grant 80NSSC21K1313, NSF Grant No. AST-2107745, and Smithsonian Astrophysical Observatory through subcontract No. SV1-21012. The AIA data are provided courtesy of NASA/SDO and the AIA science team. The SolarSoftWare (SSW) system is built from Yohkoh, SOHO, SDAC and Astronomy libraries and draws upon contributions from many members of those projects. CHIANTI is a collaborative project involving George Mason University, the University of Michigan (USA), and the University of Cambridge (UK).
\end{acknowledgments}

\appendix \label{sec:app}
\section{Density estimation for Under-intensity Flows}
We implement the approach of differential emission measure (DEM) to derive plasma density and examine the density characteristics for low intensity of the upflow, whose evolution is indicated by yellow and red arrows in Figure~\ref{evolution}), respectively. The methodology of using DEMs to derive plasma density and temperature can be found in Section 3.1 of \citet{2023ApJ...942...28X} and the references therein. The DEMs we use are based on regularized inversion that is developed by \citet{2012A&A...539A.146H} and \citet{2013A&A...553A..10H}. After obtaining DEMs, we derive the number density of electrons through 
\begin{equation}
    n = \sqrt{{\rm EM}/l}
    \label{eqn}
\end{equation} 
where 
\begin{equation}
    EM(T) = \int {\rm DEM}(T) dT,
    \label{eqem}
\end{equation}
$T$ is temperature, and $l$ is the depth of LOS with the assumption that the plasma is uniform along the LOS \citep{2017ApJ...836...55R,2020ApJ...898...88X,2023ApJ...942...28X}. Following the practice of \citet{2017ApJ...836...55R} and \citet{2020ApJ...898...88X}, we set l = 10$^{9}$~cm, the characteristic thickness of observed plasma sheets. However, we point out here that the value of $l$ does not have an impact on our intention of calculating DEMs, i.e., checking characteristics of density for low intensity in respective to the surrounding fan plasma. 

\begin{figure}[h!]
\centering
\includegraphics[scale=0.5]{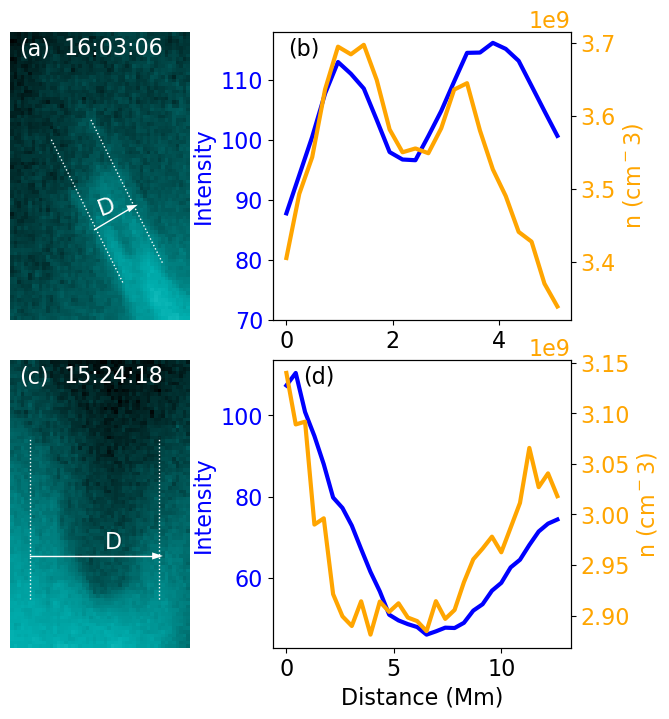}
\caption{Snapshots of upflow (a) and downflow (c) where the evolution of them can be found in Figure~\ref{evolution}. Arrows indicate the direction of distance increment. The dotted lines in Figures~\ref{ncurve} (a) and (c) indicate the paths used for averaging the values for the locations of the shortest and longest distances. Panels (b) and (d) display intensity and density n as a function of distance of the paths shown in panels (a) and (c), respectively.}
\label{ncurve}
\end{figure}

We examine the density distribution along the paths that go cross the intensity depleted upflows and downflows. The increment of distance is indicated by white arrows in Figures~\ref{evolution}(a) and (c). To enhance the signal-to-noise, for the intensity corresponding to certain distance in Figures~\ref{evolution}(b) and (d), we average the values along the paths perpendicular to the arrows. For instance, the dotted lines in Figures~\ref{ncurve} (a) and (c) indicate the paths used for averaging the values for the locations of the shortest and longest distances. The intensity and density as a function of distance are shown in Figures~\ref{evolution}(b) and (d), and it shows that intensity-depleted upflows (Figure~\ref{evolution}(c)) and downflows (Figure~\ref{evolution}(d)) are both density-depleted structures.
\section{$V$-map for brighter and darker turbulent flows}
\begin{figure}[h!]
\centering
\includegraphics[scale=0.57]{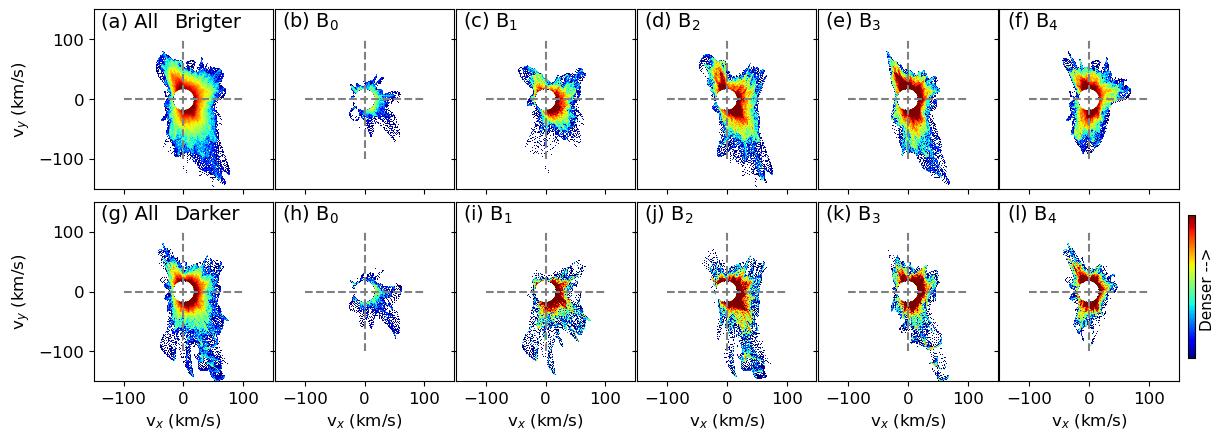}
\caption{Same as Figure~\ref{bands} but we separately plot $v$-maps for brighter (top panels) and darker (bottom panels) turbulent flows.}
\label{band-i}
\end{figure}
We further examine how darker or brighter turbulent flows contribute to characteristics of $v$-map observed at different heights. We define intensity value higher than the average value of the intensity at the height as ``brighter" and lower than the average at the height as ``darker". Figure~\ref{band-i} in the Appendix displays $v$-maps of brighter (top panels) and darker(bottom panels) turbulent flows at different heights ((b)-(f), (h)-(i)) and the whole fan area ((a) and (g)). Note that this is only for rough estimation and examine if there are distinguishable characteristics between brighter/darker turbulent flows. From Figure~\ref{band-i}, we note that, generally, there is a similar shape between brighter and darker flows of $v$-maps at each height and the whole fan area. This similarity suggests that the elongating shape (anisotropy) comes both from brighter and darker turbulent flows, suggesting the same origin of brighter and darker turbulent flows. Such characteristics have been speculated in the simulation of \citet{2022NatAs...6..317S} where MHD instabilities occur in middle and low heights in ALT, and brighter and darker turbulent flows change simultaneously. 

\section{Examination on Anisotropy of Turbulent Flows in Other Flare Events}
\label{sec:otherevents}
\begin{figure*}[ht!]
\centering
\includegraphics[scale=0.6]{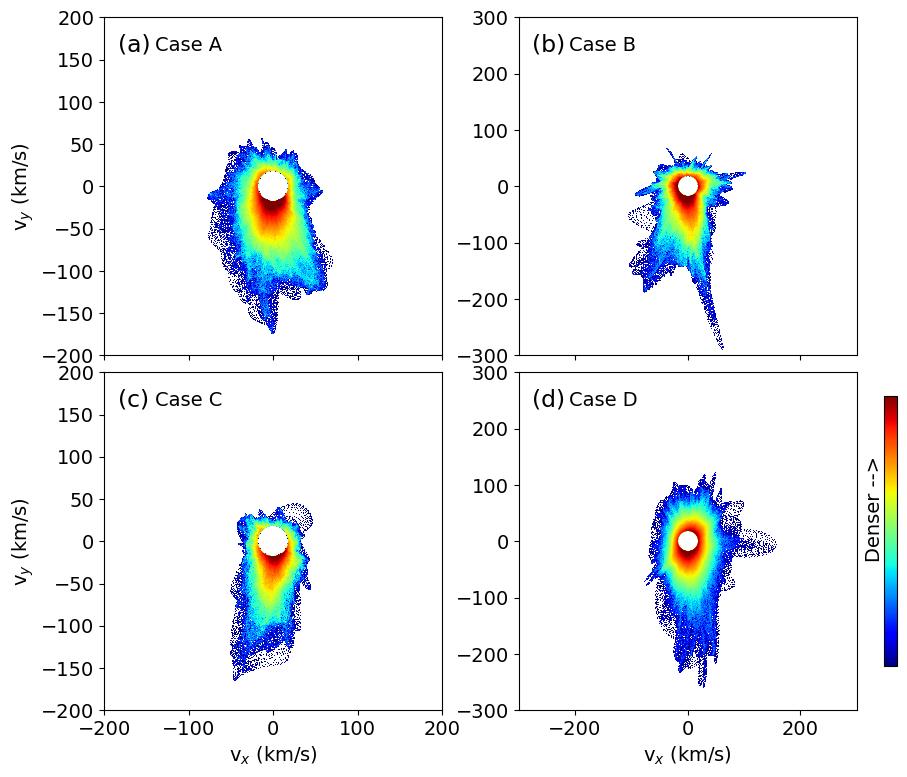}
\caption{$V$-maps for cases A-D in \citet{2022ApJ...933...15X}. The time period for panels (a) to (d) is 2012-10-22 12:02:09 -12:40:09, 2015-06-18 01:40:08-02:20:08, 2012-07-17 15:50:08-16:20:08, and 2012-01-16 04:20:09-04:50:09, respectively. In each case, we define x as the direction parallel to loop-top and y as the direction perpendicular to the loop-top. The histogram of $\theta$ in each panel is calculated by the same approach as in 2024-05-29 event.}
\label{otherevents}
\end{figure*}

Now we examine if the anisotropy along the direction perpendicular to the loop-top is a universal property in flare ALT (Figure~\ref{otherevents}). We examine cases A-D in \citet{2022ApJ...933...15X}, and the detailed description about these 4 flares can be found in \citet{2022ApJ...933...15X}. In each case, we define $x$ as the direction parallel to loop-top and $y$ as the direction perpendicular to the loop-top. The selected time period for each event features the formation of a clear supra-arcade fan on SDO/AIA 131 band observations and the manifestations of flows traveling through the fan. The $v$-maps of four examined events display consistent characteristics as in 2024-05-29 event, manifesting as the shape of $v$-maps elongates along $y$-axis direction.
\bibliography{sample631}{}

\begin{thebibliography}{}
\expandafter\ifx\csname natexlab\endcsname\relax\def\natexlab#1{#1}\fi
\providecommand{\url}[1]{\href{#1}{#1}}
\providecommand{\dodoi}[1]{doi:~\href{http://doi.org/#1}{\nolinkurl{#1}}}
\providecommand{\doeprint}[1]{\href{http://ascl.net/#1}{\nolinkurl{http://ascl.net/#1}}}
\providecommand{\doarXiv}[1]{\href{https://arxiv.org/abs/#1}{\nolinkurl{https://arxiv.org/abs/#1}}}

\bibitem[{{Asai} {et~al.}(2004){Asai}, {Yokoyama}, {Shimojo}, \& {Shibata}}]{2004ApJ...605L..77A}
{Asai}, A., {Yokoyama}, T., {Shimojo}, M., \& {Shibata}, K. 2004, \apjl, 605, L77, \dodoi{10.1086/420768}

\bibitem[{{Ashfield} {et~al.}(2024){Ashfield}, {Polito}, {Yu}, {Collier}, \& {Hayes}}]{2024ApJ...973...96A}
{Ashfield}, William, I., {Polito}, V., {Yu}, S., {Collier}, H., \& {Hayes}, L.~A. 2024, \apj, 973, 96, \dodoi{10.3847/1538-4357/ad64ca}

\bibitem[{{Chen} {et~al.}(2015){Chen}, {Bastian}, {Shen}, {Gary}, {Krucker}, \& {Glesener}}]{2015Sci...350.1238C}
{Chen}, B., {Bastian}, T.~S., {Shen}, C., {et~al.} 2015, Science, 350, 1238, \dodoi{10.1126/science.aac8467}

\bibitem[{{Chen} {et~al.}(2020){Chen}, {Shen}, {Gary}, {Reeves}, {Fleishman}, {Yu}, {Guo}, {Krucker}, {Lin}, {Nita}, \& et~al.}]{2020NatAs...4.1140C}
{Chen}, B., {Shen}, C., {Gary}, D.~E., {et~al.} 2020, Nature Astronomy, 4, 1140, \dodoi{10.1038/s41550-020-1147-7}

\bibitem[{{Chen} {et~al.}(2024){Chen}, {Kong}, {Yu}, {Shen}, {Li}, {Guo}, {Zhang}, {Glesener}, \& {Krucker}}]{Chen2024ApJ...971...85C}
{Chen}, B., {Kong}, X., {Yu}, S., {et~al.} 2024, \apj, 971, 85, \dodoi{10.3847/1538-4357/ad531a}

\bibitem[{{Chen} {et~al.}(2017){Chen}, {Liu}, {Deng}, \& {Wang}}]{2017A&A...606A..84C}
{Chen}, X., {Liu}, R., {Deng}, N., \& {Wang}, H. 2017, \aap, 606, A84, \dodoi{10.1051/0004-6361/201629893}

\bibitem[{{Cheng} {et~al.}(2018){Cheng}, {Li}, {Wan}, {Ding}, {Chen}, {Zhang}, \& {Liu}}]{2018ApJ...866...64C}
{Cheng}, X., {Li}, Y., {Wan}, L.~F., {et~al.} 2018, \apj, 866, 64, \dodoi{10.3847/1538-4357/aadd16}

\bibitem[{{Dong} {et~al.}(2018){Dong}, {Wang}, {Huang}, {Comisso}, \& {Bhattacharjee}}]{Dong2018}
{Dong}, C., {Wang}, L., {Huang}, Y.-M., {Comisso}, L., \& {Bhattacharjee}, A. 2018, PhRvL, 121, 165101, \dodoi{10.1103/PhysRevLett.121.165101}

\bibitem[{{Dong} {et~al.}(2022){Dong}, {Wang}, {Huang}, {Comisso}, {Sandstrom}, \& {Bhattacharjee}}]{Dong2022}
{Dong}, C., {Wang}, L., {Huang}, Y.-M., {et~al.} 2022, Science Advances, 8, eabn7627, \dodoi{10.1126/sciadv.abn7627}

\bibitem[{{Doschek} {et~al.}(2014){Doschek}, {McKenzie}, \& {Warren}}]{2014ApJ...788...26D}
{Doschek}, G.~A., {McKenzie}, D.~E., \& {Warren}, H.~P. 2014, \apj, 788, 26, \dodoi{10.1088/0004-637X/788/1/26}

\bibitem[{Farneb{\"a}ck(2003)}]{10.1007/3-540-45103-X_50}
Farneb{\"a}ck, G. 2003, in Image Analysis, ed. J.~Bigun \& T.~Gustavsson (Berlin, Heidelberg: Springer Berlin Heidelberg), 363--370

\bibitem[{{Fleishman} {et~al.}(2022){Fleishman}, {Nita}, {Chen}, {Yu}, \& {Gary}}]{2022Natur.606..674F}
{Fleishman}, G.~D., {Nita}, G.~M., {Chen}, B., {Yu}, S., \& {Gary}, D.~E. 2022, \nat, 606, 674, \dodoi{10.1038/s41586-022-04728-8}

\bibitem[{{Forbes}(2000)}]{2000JGR...10523153F}
{Forbes}, T.~G. 2000, \jgr, 105, 23153, \dodoi{10.1029/2000JA000005}

\bibitem[{{Freed} \& {McKenzie}(2018)}]{2018ApJ...866...29F}
{Freed}, M.~S., \& {McKenzie}, D.~E. 2018, \apj, 866, 29, \dodoi{10.3847/1538-4357/aadee4}

\bibitem[{{Guo} {et~al.}(2014){Guo}, {Huang}, {Bhattacharjee}, \& {Innes}}]{2014ApJ...796L..29G}
{Guo}, L.~J., {Huang}, Y.~M., {Bhattacharjee}, A., \& {Innes}, D.~E. 2014, \apjl, 796, L29, \dodoi{10.1088/2041-8205/796/2/L29}

\bibitem[{{Hannah} \& {Kontar}(2012)}]{2012A&A...539A.146H}
{Hannah}, I.~G., \& {Kontar}, E.~P. 2012, \aap, 539, A146, \dodoi{10.1051/0004-6361/201117576}

\bibitem[{{Hannah} \& {Kontar}(2013)}]{2013A&A...553A..10H}
---. 2013, \aap, 553, A10, \dodoi{10.1051/0004-6361/201219727}

\bibitem[{{Innes} {et~al.}(2014){Innes}, {Guo}, {Bhattacharjee}, {Huang}, \& {Schmit}}]{2014ApJ...796...27I}
{Innes}, D.~E., {Guo}, L.~J., {Bhattacharjee}, A., {Huang}, Y.~M., \& {Schmit}, D. 2014, \apj, 796, 27, \dodoi{10.1088/0004-637X/796/1/27}

\bibitem[{{Innes} {et~al.}(2003){Innes}, {McKenzie}, \& {Wang}}]{2003SoPh..217..247I}
{Innes}, D.~E., {McKenzie}, D.~E., \& {Wang}, T. 2003, \solphys, 217, 247, \dodoi{10.1023/B:SOLA.0000006899.12788.22}

\bibitem[{{Kim} {et~al.}(2014){Kim}, {Shibasaki}, {Bain}, \& {Cho}}]{2014ApJ...785..106K}
{Kim}, S., {Shibasaki}, K., {Bain}, H.~M., \& {Cho}, K.~S. 2014, \apj, 785, 106, \dodoi{10.1088/0004-637X/785/2/106}

\bibitem[{{Kong} {et~al.}(2020){Kong}, {Guo}, {Shen}, {Chen}, {Chen}, \& {Giacalone}}]{2020ApJ...905L..16K}
{Kong}, X., {Guo}, F., {Shen}, C., {et~al.} 2020, \apjl, 905, L16, \dodoi{10.3847/2041-8213/abcbf5}

\bibitem[{{Kong} {et~al.}(2022){Kong}, {Chen}, {Guo}, {Shen}, {Li}, {Ye}, {Zhao}, {Jiang}, {Yu}, {Chen}, \& {Giacalone}}]{Kong2022ApJ...941L..22K}
{Kong}, X., {Chen}, B., {Guo}, F., {et~al.} 2022, \apjl, 941, L22, \dodoi{10.3847/2041-8213/aca65c}

\bibitem[{{Kontar} {et~al.}(2017){Kontar}, {Perez}, {Harra}, {Kuznetsov}, {Emslie}, {Jeffrey}, {Bian}, \& {Dennis}}]{2017PhRvL.118o5101K}
{Kontar}, E.~P., {Perez}, J.~E., {Harra}, L.~K., {et~al.} 2017, \prl, 118, 155101, \dodoi{10.1103/PhysRevLett.118.155101}

\bibitem[{{Li} {et~al.}(2022){Li}, {Guo}, {Chen}, {Shen}, \& {Glesener}}]{2022ApJ...932...92L}
{Li}, X., {Guo}, F., {Chen}, B., {Shen}, C., \& {Glesener}, L. 2022, \apj, 932, 92, \dodoi{10.3847/1538-4357/ac6efe}

\bibitem[{{Li} {et~al.}(2019){Li}, {Guo}, {Li}, {Stanier}, \& {Kilian}}]{Li2019ApJ...884..118L}
{Li}, X., {Guo}, F., {Li}, H., {Stanier}, A., \& {Kilian}, P. 2019, \apj, 884, 118, \dodoi{10.3847/1538-4357/ab4268}

\bibitem[{{Li} {et~al.}(2021){Li}, {Cheng}, {Ding}, {Reeves}, {Kittrell}, {Weber}, \& {McKenzie}}]{2021ApJ...915..124L}
{Li}, Z.~F., {Cheng}, X., {Ding}, M.~D., {et~al.} 2021, \apj, 915, 124, \dodoi{10.3847/1538-4357/ac043e}

\bibitem[{{Liu} \& {Wang}(2021)}]{2021A&A...653A..51L}
{Liu}, R., \& {Wang}, Y. 2021, \aap, 653, A51, \dodoi{10.1051/0004-6361/202140847}

\bibitem[{{Mann} {et~al.}(2009){Mann}, {Warmuth}, \& {Aurass}}]{2009A&A...494..669M}
{Mann}, G., {Warmuth}, A., \& {Aurass}, H. 2009, \aap, 494, 669, \dodoi{10.1051/0004-6361:200810099}

\bibitem[{{McKenzie}(2000)}]{2000SoPh..195..381M}
{McKenzie}, D.~E. 2000, \solphys, 195, 381, \dodoi{10.1023/A:1005220604894}

\bibitem[{{McKenzie}(2013)}]{2013ApJ...766...39M}
---. 2013, \apj, 766, 39, \dodoi{10.1088/0004-637X/766/1/39}

\bibitem[{{McKenzie} \& {Hudson}(1999)}]{1999ApJ...519L..93M}
{McKenzie}, D.~E., \& {Hudson}, H.~S. 1999, \apjl, 519, L93, \dodoi{10.1086/312110}

\bibitem[{{McKenzie} \& {Savage}(2009)}]{2009ApJ...697.1569M}
{McKenzie}, D.~E., \& {Savage}, S.~L. 2009, \apj, 697, 1569, \dodoi{10.1088/0004-637X/697/2/1569}

\bibitem[{{McKenzie} \& {Savage}(2011)}]{2011ApJ...735L...6M}
---. 2011, \apjl, 735, L6, \dodoi{10.1088/2041-8205/735/1/L6}

\bibitem[{{Mei} {et~al.}(2017){Mei}, {Keppens}, {Roussev}, \& {Lin}}]{2017A&A...604L...7M}
{Mei}, Z.~X., {Keppens}, R., {Roussev}, I.~I., \& {Lin}, J. 2017, \aap, 604, L7, \dodoi{10.1051/0004-6361/201731146}

\bibitem[{{Milligan}(2011)}]{2011ApJ...740...70M}
{Milligan}, R.~O. 2011, \apj, 740, 70, \dodoi{10.1088/0004-637X/740/2/70}

\bibitem[{{Polito} {et~al.}(2018){Polito}, {Dud{\'\i}k}, {Ka{\v{s}}parov{\'a}}, {Dzif{\v{c}}{\'a}kov{\'a}}, {Reeves}, {Testa}, \& {Chen}}]{2018ApJ...864...63P}
{Polito}, V., {Dud{\'\i}k}, J., {Ka{\v{s}}parov{\'a}}, J., {et~al.} 2018, \apj, 864, 63, \dodoi{10.3847/1538-4357/aad62d}

\bibitem[{{Reeves} {et~al.}(2017){Reeves}, {Freed}, {McKenzie}, \& {Savage}}]{2017ApJ...836...55R}
{Reeves}, K.~K., {Freed}, M.~S., {McKenzie}, D.~E., \& {Savage}, S.~L. 2017, \apj, 836, 55, \dodoi{10.3847/1538-4357/836/1/55}

\bibitem[{{Reeves} \& {Golub}(2011)}]{2011ApJ...727L..52R}
{Reeves}, K.~K., \& {Golub}, L. 2011, \apjl, 727, L52, \dodoi{10.1088/2041-8205/727/2/L52}

\bibitem[{{Reeves} {et~al.}(2020){Reeves}, {Polito}, {Chen}, {Galan}, {Yu}, {Liu}, \& {Li}}]{2020ApJ...905..165R}
{Reeves}, K.~K., {Polito}, V., {Chen}, B., {et~al.} 2020, \apj, 905, 165, \dodoi{10.3847/1538-4357/abc4e0}

\bibitem[{{Ruan} {et~al.}(2023){Ruan}, {Yan}, \& {Keppens}}]{2023ApJ...947...67R}
{Ruan}, W., {Yan}, L., \& {Keppens}, R. 2023, \apj, 947, 67, \dodoi{10.3847/1538-4357/ac9b4e}

\bibitem[{{Samanta} {et~al.}(2021){Samanta}, {Tian}, {Chen}, {Reeves}, {Cheung}, {Vourlidas}, \& {Banerjee}}]{2021Innov...200083S}
{Samanta}, T., {Tian}, H., {Chen}, B., {et~al.} 2021, The Innovation, 2, 100083, \dodoi{10.1016/j.xinn.2021.100083}

\bibitem[{{Samanta} {et~al.}(2019){Samanta}, {Tian}, \& {Nakariakov}}]{2019PhRvL.123c5102S}
{Samanta}, T., {Tian}, H., \& {Nakariakov}, V.~M. 2019, \prl, 123, 035102, \dodoi{10.1103/PhysRevLett.123.035102}

\bibitem[{{Savage} \& {McKenzie}(2011)}]{2011ApJ...730...98S}
{Savage}, S.~L., \& {McKenzie}, D.~E. 2011, \apj, 730, 98, \dodoi{10.1088/0004-637X/730/2/98}

\bibitem[{{Savage} {et~al.}(2012){Savage}, {McKenzie}, \& {Reeves}}]{2012ApJ...747L..40S}
{Savage}, S.~L., {McKenzie}, D.~E., \& {Reeves}, K.~K. 2012, \apjl, 747, L40, \dodoi{10.1088/2041-8205/747/2/L40}

\bibitem[{{Savage} {et~al.}(2010){Savage}, {McKenzie}, {Reeves}, {Forbes}, \& {Longcope}}]{2010ApJ...722..329S}
{Savage}, S.~L., {McKenzie}, D.~E., {Reeves}, K.~K., {Forbes}, T.~G., \& {Longcope}, D.~W. 2010, \apj, 722, 329, \dodoi{10.1088/0004-637X/722/1/329}

\bibitem[{{Shen} {et~al.}(2022){Shen}, {Chen}, {Reeves}, {Yu}, {Polito}, \& {Xie}}]{2022NatAs...6..317S}
{Shen}, C., {Chen}, B., {Reeves}, K.~K., {et~al.} 2022, Nature Astronomy, 6, 317, \dodoi{10.1038/s41550-021-01570-2}

\bibitem[{{Shen} {et~al.}(2018){Shen}, {Kong}, {Guo}, {Raymond}, \& {Chen}}]{2018ApJ...869..116S}
{Shen}, C., {Kong}, X., {Guo}, F., {Raymond}, J.~C., \& {Chen}, B. 2018, \apj, 869, 116, \dodoi{10.3847/1538-4357/aaeed3}

\bibitem[{{Shibata} {et~al.}(2023){Shibata}, {Takasao}, \& {Reeves}}]{2023ApJ...943..106S}
{Shibata}, K., {Takasao}, S., \& {Reeves}, K.~K. 2023, \apj, 943, 106, \dodoi{10.3847/1538-4357/acaa9c}

\bibitem[{{Takahashi} {et~al.}(2017){Takahashi}, {Qiu}, \& {Shibata}}]{2017ApJ...848..102T}
{Takahashi}, T., {Qiu}, J., \& {Shibata}, K. 2017, \apj, 848, 102, \dodoi{10.3847/1538-4357/aa8f97}

\bibitem[{{Takasao} \& {Shibata}(2016)}]{2016ApJ...823..150T}
{Takasao}, S., \& {Shibata}, K. 2016, \apj, 823, 150, \dodoi{10.3847/0004-637X/823/2/150}

\bibitem[{{Tan} {et~al.}(2022){Tan}, {Hou}, \& {Tian}}]{2022MNRAS.516.3120T}
{Tan}, G., {Hou}, Y., \& {Tian}, H. 2022, \mnras, 516, 3120, \dodoi{10.1093/mnras/stac2470}

\bibitem[{{Wang} {et~al.}(2022){Wang}, {Cheng}, {Ren}, \& {Ding}}]{2022ApJ...931L..32W}
{Wang}, Y., {Cheng}, X., {Ren}, Z., \& {Ding}, M. 2022, \apjl, 931, L32, \dodoi{10.3847/2041-8213/ac715a}

\bibitem[{{Warren} {et~al.}(2018){Warren}, {Brooks}, {Ugarte-Urra}, {Reep}, {Crump}, \& {Doschek}}]{2018ApJ...854..122W}
{Warren}, H.~P., {Brooks}, D.~H., {Ugarte-Urra}, I., {et~al.} 2018, \apj, 854, 122, \dodoi{10.3847/1538-4357/aaa9b8}

\bibitem[{{Warren} {et~al.}(2011){Warren}, {O'Brien}, \& {Sheeley}}]{2011ApJ...742...92W}
{Warren}, H.~P., {O'Brien}, C.~M., \& {Sheeley}, Neil~R., J. 2011, \apj, 742, 92, \dodoi{10.1088/0004-637X/742/2/92}

\bibitem[{{Xie} {et~al.}(2024){Xie}, {Li}, {Reeves}, \& {Gou}}]{2024FrASS..1183746X}
{Xie}, X., {Li}, G., {Reeves}, K.~K., \& {Gou}, T. 2024, Frontiers in Astronomy and Space Sciences, 11, 1383746, \dodoi{10.3389/fspas.2024.1383746}

\bibitem[{{Xie} {et~al.}(2022{\natexlab{a}}){Xie}, {Mei}, {Shen}, {Cai}, {Ye}, {Reeves}, {Roussev}, \& {Lin}}]{2022MNRAS.509..406X}
{Xie}, X., {Mei}, Z., {Shen}, C., {et~al.} 2022{\natexlab{a}}, \mnras, 509, 406, \dodoi{10.1093/mnras/stab2954}

\bibitem[{{Xie} \& {Reeves}(2023)}]{2023ApJ...942...28X}
{Xie}, X., \& {Reeves}, K.~K. 2023, \apj, 942, 28, \dodoi{10.3847/1538-4357/ac9f47}

\bibitem[{{Xie} {et~al.}(2022{\natexlab{b}}){Xie}, {Reeves}, {Shen}, \& {Ingram}}]{2022ApJ...933...15X}
{Xie}, X., {Reeves}, K.~K., {Shen}, C., \& {Ingram}, J.~D. 2022{\natexlab{b}}, \apj, 933, 15, \dodoi{10.3847/1538-4357/ac695d}

\bibitem[{{Xue} {et~al.}(2020){Xue}, {Su}, {Li}, \& {Zhao}}]{2020ApJ...898...88X}
{Xue}, J., {Su}, Y., {Li}, H., \& {Zhao}, X. 2020, \apj, 898, 88, \dodoi{10.3847/1538-4357/ab9a3d}

\bibitem[{{Ye} {et~al.}(2023){Ye}, {Raymond}, {Mei}, {Cai}, {Chen}, {Li}, \& {Lin}}]{2023ApJ...955...88Y}
{Ye}, J., {Raymond}, J.~C., {Mei}, Z., {et~al.} 2023, \apj, 955, 88, \dodoi{10.3847/1538-4357/acf129}

\bibitem[{{Yu} {et~al.}(2020){Yu}, {Chen}, {Reeves}, {Gary}, {Musset}, {Fleishman}, {Nita}, \& {Glesener}}]{2020ApJ...900...17Y}
{Yu}, S., {Chen}, B., {Reeves}, K.~K., {et~al.} 2020, \apj, 900, 17, \dodoi{10.3847/1538-4357/aba8a6}

\bibitem[{{Zhang} {et~al.}(2019){Zhang}, {Buchlin}, \& {Vial}}]{2019A&A...624A..72Z}
{Zhang}, P., {Buchlin}, {\'E}., \& {Vial}, J.~C. 2019, \aap, 624, A72, \dodoi{10.1051/0004-6361/201834259}

\end{thebibliography}
\bibliographystyle{aasjournal}



\end{document}